\documentclass[
aps,
11pt,
final,
notitlepage,
oneside,
nobibnotes,
nofootinbib,
superscriptaddress,
showpacs]
{revtex4}
\usepackage{graphicx}
\usepackage{amsmath}
\usepackage{amsfonts}
\usepackage{amssymb}
\usepackage{bbm}  % blackboard math
\usepackage{young}  % Young diagrams
\usepackage{yfonts}  % Fraktur für Complexe
\newcommand{\MO}{\protect{M{\"o}bius}}
\newcommand{\PL}{\protect{Pl{\"u}cker}}
\newcommand{\PLL}{\protect{Pl{\"u}cker }}
\begin{document}
\title{On a Microscopic Representation of Space-Time V}
\author{\firstname{Rolf}~\surname{Dahm}}
%\email{dahm@bf-IS.de}
\affiliation{Permanent address: beratung f{\"{u}}r Informationssysteme und Systemintegration,
G{\"{a}}rtnergasse 1, D-55116 Mainz, Germany}

\begin{abstract}
In order to extend our approach based on SU$*$(4), we were led to 
(real) projective and (line) Complex geometry. So here we start 
from quadratic Complexe which yield naturally the 'light cone' 
$x_{1}^{2}+x_{2}^{2}+x_{3}^{2}-x_{0}^{2}=0$ when being related to 
(homogeneous) point coordinates $x_{\alpha}^{2}$ and infinitesimal 
dynamics by tetrahedral Complexe (or line elements). This introduces 
naturally projective transformations by preserving anharmonic ratios. 
Referring to old work of Pl{\"u}cker relating quadratic Complexe to 
optics, we discuss (linear) symplectic symmetry and line coordinates, 
the main purpose and thread within this paper, however, is the 
identification and discussion of special relativity as direct 
invariance properties of line/Complex coordinates as well as 
their relation to 'quantum field theory' by complexification of
point coordinates or Complexe. This can be established by the 
Lie mapping which relates lines/Complexe to sphere geometry so 
that SU(2), SU(2)$\times$U(1), SU(2)$\times$SU(2) and the Dirac 
spinor description emerge without additional assumptions. We give 
a short outlook in that quadratic Complexe are related to dynamics 
e.g.~power expressions in terms of six-vector products of Complexe, 
and action principles may be applied. (Quadratic) products like 
$F^{\mu\nu}F_{\mu\nu}$ or $F^{a\,\mu\nu}F^{a}_{\mu\nu}$, $1\leq a\leq 3$ 
are natural quadratic Complex expressions ('invariants') which may 
be extended by line constraints $\lambda k\cdot\epsilon=0$ with 
respect to an 'action principle' so that we identify 'quantum field 
theory' with projective or line/Complex geometry having applied the 
Lie mapping.
\end{abstract}


\maketitle

\pagebreak

\section{Introduction}
With this fifth part\footnote{This paper, being limited as conference
proceedings, yields mainly the direct arguments and calculations related
to the abstract. A more detailed paper containing additional background
and details is going to appear as part VI soon.} of our series, we've 
reached a certain milestone to pause and make up the balance in departing
from SU(4)~vs.~SU$*$(4), the various derived symmetry breaking patterns 
\cite{dahm:2008}, \cite{dahm:MRST1} and the identification of photons in
SSB patterns. So starting from the (physical!) interpretation of the Lie
algebra su(4) (in terms of spin, isospin and chiral symmetry) and its 
various real forms, the main path -- also in the context of noncompact 
su$*$(4) and the Dirac algebra -- concentrated on a 10$\oplus$5 reductive
decomposition and its interpretation (see \cite{dahm:MRST1} and references).
In collecting various algebraic and physical aspects, we were led to 
projective geometry, transfer principles and especially line and Complex
geometry \cite{dahm:MRST3}, \cite{dahm:MRST4}.

Here, we want to present briefly some additional details and identifications
with respect to typical/basic 'ingredients' of dynamical theories like 
special relativity, Dirac's 'square root' of the Klein-Gordon equation,
and -- most important -- identification of states resp. interpretation 
of the mathematical description. Whereas a 'light cone', Lorentz transformations
and special relativity can be related to (real) projective geometry, 
we need the Lie mapping \cite{lie:1872} to introduce Pauli matrices
(respectively quaternions) which relates to 'quantum theory' and 
'quantum' notion.

The central theme is\footnote{Here, we discuss 'simple' geometrical issues 
and identifications according to \protect{\cite{lie:1872}} and postpone 
the viewpoint of 'advanced geometry' in terms of the \PL-Klein quadric in 
\protect{$\mathbbm{R}^5$}, Study's points of view \protect{\cite{study:1903}}, 
or even higher dimensional transfers to upcoming publications/discussions.}
that this transformation relates {\it line} representations\footnote{As 
before, subsequently we use the shorthand notation 'rep'.} (NOT {\it point} 
reps!) to spheres (and elliptic geometry). So first of all we have to use 
and respect a priori {\it two} distinct spaces (and not one and the same 
space-time) respectively their related individual 'physical' identification(s)
and, more important, our fundamental geometrical space element is a (nonlocal) 
4-dim {\it line} and afterwards a set of lines, the linear Complex, not as 
usual the point or 'vector' rep. So points emerge only as line incidences
which must be treated by projective geometry. However, dynamics being 
concerned with tangents, momenta and conic sections benefits ernormously 
according to this different choice of the fundamental geometrical space 
element due to (linear) Complexe being related to null systems, the 
Hamilton formalism and differential equations \cite{plueckerNG:1868},
\cite{lie:1896}. This comprises a very long history where lines, various
compositions of lines (i.e.~Complexe, congruences, etc.) and null systems
have been applied with great use and success to physics. So here, we start
right from the beginning with line and Complex geometry\footnote{For 
introductory details from physical identifications to projective geometry, 
please see \protect{\cite{dahm:MRST3}} and \protect{\cite{dahm:MRST4}}, 
some basic definitions have been repeated in \protect{\cite{dahm:MRST3}} 
section 1.2ff which originate mainly from \PL's work \cite{plueckerNG:1868}
on line Complexe.}.

\section{Quadratic Complexe and the 'Light Cone'}
% Here, we want to extend this approach further in terms of line and Complex 
% geometry and give some analytical examples. As such, we start from quadratic 
% Complexe which we've identified in part IV already as yielding naturally the 
% 'light cone' $x_{1}^{2}+x_{2}^{2}+x_{3}^{2}-x_{0}^{2}=0$ while being related 
Why do we want to see the 'light cone' $x_{1}^{2}+x_{2}^{2}+x_{3}^{2}-x_{0}^{2}=0$
(in point coordinates) emerging directly from quadratic Complexe like 
in \protect{\cite{dahm:MRST3}}? Well, the reason is simple: As soon 
as we interpret this 'light cone' (as usual) in terms of a 'metric' 
on point spaces and/or in metric coordinates, we have to introduce 
additional physical identification or at least an additional dimension
(i.e. we have to use {\it five} homogeneous coordinates, see e.g.~\cite{klein:1928},
appendix~\S 5) in order to treat absolute elements ('infinities'). 
From our viewpoint, it is much easier and much more consistent (and 
we'll give some additional arguments below) to understand the 'light 
cone' as an absolute element (or 'gauge surface') when switching 
from quadratic line/Complex reps to point reps. This allows to introduce
and relate velocities via a 'common' (system) time parameter $t$ to
point coordinates $x_{\alpha}$, i.e.~the geometrical line picture to
a dynamic point picture and as such to relations/ratios of velocities
in terms of $\beta_{i}=v_{i}/c$. So the velocity of light $c$ may serve
to parametrize the absolute coordinate $x_{0}$ by its correspondence
to physical observations and identifications, using the velocity picture
$x_{i}\sim v_{i}t$, of course (see \protect{\cite{dahm:MRST3}}), and 
the 'gauge surface' in the geometrical line picture to apply the 
Cayley-Klein mechanism towards metric coordinates and spaces. We 
have mentioned already the introduction of equivalence classes by
this velocity concept. Although it is known from classical mechanics
how to work with (euclidean) coordinate projections and 'vectors',
it is also known\footnote{As an example, we cite \protect{\cite{schoenflies:1884}} 
which, starting from Complex description of rigid bodies, gives 
insight into some deeper aspects of 'advanced geometry' by addressing
null systems and the tetrahedral Complex, 3rd and higher order 
elements, etc.} that in order to describe kinematics and dynamics,
the naive (3-dim euclidean) 'vector' picture of velocities, momenta
and forces is not sufficient but one has to work with force and 
null systems which were generalized by \PLL to Complexe and 6-dim
'Dynamen' \cite{pluecker:1866}. This means, we use {\it four 
homogeneous} coordinates $x_{\alpha}$, and the 'metric'/the 
'light cone' is a point representation related to the line 
representation of a quadratic Complex. So this picture yields
a natural relation in the context of projective geometry of 
3-space, and natural representations of lines in terms of 
points and planes (see \protect{\cite{dahm:MRST3}}, p.~9, 
and references).

Here, we skip some background on quadratic Complexe and postpone
the details to part VI. In brief, having mentioned (\cite{dahm:MRST4},
III C) \PL's work on relating ellipsoids and lines/Complexe already,
some more aspects can be found in Klein's book\footnote{For the sake
of simplicity, we use this book and the references given there as a
kind of dictionary in order to avoid lengthy details or historical
background.} \cite{klein:1926} \S\S 4-7 with respect to potential
theory, confocal surfaces and elliptic coordinates. Physical 
applications in optics \cite{pluecker:1865}, \cite{plueckerNG:1868}
should be mentioned, too, in that double refraction at a crystal
surface yields a triple (line) vertex or in generalizing the 
variational action principle \cite{pluecker:1866}, \cite{klein:1871}.

In addition, we have learned from \cite{ehlers:1972} a basic set
of requirements to express relativity, and we have to state that
-- being in charge to introduce those requirements in point reps
and differential geometry -- it is Complex geometry (and especially
in conjunction with quadratic Complexe) which yields the requirements
with respect to necessary conformal, projective and affine structures
automatically, especially such non-local requirements like second-order
cones at each point of the curve or geodesic families with certain
behavior. So using quadratic Complexe, we control a superset {\it to 
derive} those features -- there is no need to introduce them by hand,
but we have to use Complex geometry. So the unifying space element
is a linear Complex, and we have to relate our reasoning in 3-dim 
space to higher order Complexe and calculation patterns in order 
to compare to physics and extract principles.

\subsection{Infinitesimal Dynamics}
% maybe more on quadratic Complexe, i.e. tangential Complex?
% (reasoning: quadratic tetrahedral Complex interpreted as tangential Complex
%  leads to Ordnungsfläche zweiten Grades and (thus naturally) to LT)
Recalling tetrahedral Complexe, those quadratic Complexe (while preserving
by definition anharmonic ratios, see \protect{\cite{dahm:MRST4}} and 
references there) emphasize projective transformation, i.e. transformations
leaving the anharmonic ratio invariant. This provides the most 'natural link'
via Klein's Erlanger program to introduce and justify group theory in terms
of linear reps. So departing here, we may study projective transformations
in various reps e.g.~only on appropriate point sets, their transformation
groups or even only infinitesimal transformations related to them\footnote{We
want to remember that projective geometry of \protect{$\mathbbm{R}^3$} not 
only involves 'collineations' but also 'correlations' and dual elements, 
and it allows for application of higher transfer principles! In the context
of transfer principles, it is necessary to recall the identification of 
certain {\it point} coordinates in higher dimensional spaces with (even 
extended) geometrical elements in ordinary real 3-dim space, e.g.~points 
with respect to the \PL-Klein quadric or the Lie mapping later in 
section~\protect{\ref{ch:liemapping}}.}. With respect to more general
geometric considerations, however, by changing the underlying space
element (e.g.~from points to lines or spheres), we want to make first
steps into linear and quadratic Complex geometry\footnote{Of course,
we may choose other elements and transfer them to (higher-dim) point
spaces. Here, we stay in the regime of 3-dim space, the Complex as
basic element, the Lie mapping, sphere and Laguerre geometry.}.
Additional details can be found in \cite{klein:1926} \S\S 21-24.

Now the shorthand approach to dynamics can be based on line incidence
and in a second step on Complex involutions \cite{klein:1871}. Whereas
the incidence relation of two lines with coordinates $p_{\alpha\beta}$
and $p'_{\alpha\beta}$ reads as (\cite{klein:1926} \S 20)
\begin{equation}
\label{eq:lineincidence}
p_{12}p'_{34}+p_{13}p'_{42}+p_{14}p'_{23}
+p_{34}p'_{12}+p_{42}p'_{13}+p_{23}p'_{14}=0\,,
\end{equation} 
choosing the linear Complex as basic element introduces new notations
and invariant structures \cite{klein:1926} \S\S 21-24. The incidence 
equation (\ref{eq:lineincidence}) above, with line coordinates $p_{\alpha\beta}$
being replaced by Complex coordinates, declare involution of Complexe,
and in analyzing the quadratic form $\Omega$, Klein introduced \cite{klein:1869},
\cite{klein:1871} six fundamental Complexe and determined their invariants
to be $\pm 1$. Because of their relation to handedness, the six fundamental
Complexe\footnote{With respect to \protect{SU(2)$\times$SU(2)} structures,
we still have to introduce the Lie mapping, see section~\protect{\ref{ch:liemapping}}.}
decompose into $3\oplus 3$ left-/right-handed, the simultaneous invariant
of each pair being 0.

If -- as before \cite{dahm:MRST3} -- we identify $F^{\mu\nu}F_{\mu\nu}$
as a Complex invariant, on the one hand we can use Klein's work to
postulate an action (see section \ref{ch:energy}), on the other hand
one has to recall (see e.g.~\cite{moebius:1837}, \cite{pluecker:1866},
\cite{klein:1869} or \cite{klein:1871}) the discussion of up to six
acting forces. So with respect to $F^{\mu\nu}F_{\mu\nu}$ and 
$F^{a\,\mu\nu}F^{a}_{\mu\nu}$, $1\leq a\leq 3$, emerging in actions,
we have settled the basic environment here from within the framework
of Complex geometry and involutions.

Last not least, it is noteworthy that the (infinitesimal) action (and as
such the very foundation of the action principle) may be written as a 
quadratic Complex (see e.g.~Klein \cite{klein:1928} or Dirac \cite{dirac:1973PRS}~eq.~(5.1))
so that projective transformations are automatically transformation groups
of the action (which of course allows to apply Klein's Erlanger program in
all details with respect to (linear) transformation groups and subgroups).
We've had related $F^{\mu\nu}$ already to a (linear) Complex \cite{dahm:MRST3},
so starting from the known rep $F^{\mu\nu}$, $\Omega$ in \cite{dirac:1973PRS},~eq.~(5.1)
suggests several interpretations and generalizations. Thus $\Omega$ can
be interpreted in terms of incidence/conjugation of lines, involution of 
linear Complexe or generalized in terms of squares of two (general linear)
Complexe respectively the most general possible form of a quadratic Complex.
Moreover, this is not unique in that \cite{dirac:1973PRS},~eq.~(5.1) allows
to add other Complex reps as well as additional (incidence/conjugation) terms
equating to 0 as long as the quadratic character of $\Omega$ is preserved\footnote{In
order to preserve the grade of a Complex under projective transformations, one 
has to use \PL's 5-dim (inhomogeneous) coordinates $r$, $s$, $\rho$, $\sigma$,
and $\eta=r\sigma-s\rho$, or six homogeneous coordinates, {\it not} only 
the original 4-dim set of line coordinates $r$, $s$, $\rho$ and $\sigma$!
This introduces the (quadratic) \PLL condition which may be used as an
additional term while introducing an additional parameter into the action.
Please note here for later use, that if we introduce a matrix notation
\protect{$M:=
\left(
\begin{array}{cc}
r & \rho \\ s & \sigma
\end{array}
\right)$} then \protect{$\eta=\det M$}. So at least for linear transformations
of line coordinates this matrix rep provides a calculus (or symbolism)
to transform line coordinates by performing matrix transformations.}.

\section{Complexe and Symplectic Symmetry}
The notion 'symplectic symmetry' emerged because Weyl renamed the 
symmetry of the 'Komplexgruppe' to greek notion in order to avoid
misconceptions and/or misunderstanding of complex numbers and
Complexe\footnote{We hope to avoid such misery by strictly using 
capitalized 'C' and old german plural form in conjunction with the
treatment of line Complexe.}. We use linear Complexe and their
point rep as given e.g.~by Hamermesh (see \cite{hamermesh:1962},
ch.~10, eq.~(10-73)).

\subsection{(Linear) Symplectic Symmetry}
% Hamermesh and symplectic group in terms of point coordinates
If we do {\it not} split $n=2\nu$ and the given decomposition into 
$(n-2)$-dim subspaces \cite{hamermesh:1962}, but if instead we use
general $n$ and fix $n=4$ in a linear enumeration of the coordinate
reps $x$ and $y$, we may write the bilinear form according to
$x^{T}Jy=x^{T}_{\alpha}J_{\alpha\beta}y_{\beta}$, $1\leq\alpha,\beta\leq 4$,
which (in this special case) leads to
$x_{1}y_{2}-x_{2}y_{1}+x_{3}y_{4}-x_{4}y_{3}$. Interpreting the
coordinates as homogeneous coordinates, we may replace them by 
% identification of line coordinates, i.e. p_{12}\pm p_{34}
line coordinate reps (see e.g.~\cite{dahm:MRST3}, ch.~1.2), i.e.
the invariance requirement of $x^{T}Jy$ using point coordinates
reads as the invariance of $p_{12}+p_{34}$ in terms of (homogeneous!)
line coordinates $p_{\alpha\beta}$. Now, $p_{12}+p_{34}$ is a special
case of a general linear Complex $A_{\alpha\beta}p_{\alpha\beta}$,
and due to simplicity we'll use the expression $p_{12}\pm k p_{34}=0$
for subsequent discussions, $k\in\mathbbm{R}$ (for the moment) being
nothing but a real parameter.

\subsection{Special Relativity as Invariance Property}
% special relativity as invariance of p_{12}\pm p_{34}
Having established the (symplectic) invariance requirement of 
$x^{T}Jy=p_{12}+p_{34}$ and generalized it to $p_{12}\pm k p_{34}=0$,
we may ask for (point coordinate) transformations keeping such
objects invariant. In selecting the two most trivial cases\footnote{We
can forget all background and think with respect to \protect{$p_{\alpha\beta}$}
just in terms of \protect{$2\times 2$} determinants of some point coordinates.},
it is obvious that at a first glance, we can keep $p_{\alpha\beta}$
invariant by itself when keeping the coordinates (and as such also
the area element) $x_{\alpha},y_{\alpha}\longrightarrow x'_{\alpha}=x_{\alpha},y'_{\alpha}=y_{\alpha}$
invariant.

A second, almost trivial possibility is to apply transformations
according to
\begin{equation}
\label{eq:lt}
\begin{array}{cc}
x_{1}\longrightarrow x'_{1}=x_{1}\,,\quad x_{2}\longrightarrow x'_{2}=x_{2}\,,\quad &
y_{1}\longrightarrow y'_{1}=y_{1}\,,\quad y_{2}\longrightarrow y'_{2}=y_{2}\,,\\
x_{3}\longrightarrow x'_{3}=x_{3}\,\cosh\eta-x_{4}\,\sinh\eta\,,\quad &
y_{3}\longrightarrow y'_{3}=y_{3}\,\cosh\eta-y_{4}\,\sinh\eta\,, \\
x_{4}\longrightarrow x'_{4}=x_{4}\,\cosh\eta-x_{3}\,\sinh\eta\,,\quad &
y_{4}\longrightarrow x'_{4}=y_{4}\,\cosh\eta-y_{3}\,\sinh\eta\,.
\end{array}
\end{equation}
Direct calculation shows $(x'_{3})^{2}-(x'_{4})^{2}=x_{3}^{2}-x_{4}^{2}$, or
even $(x'_{1})^{2}+(x'_{2})^{2}+(x'_{3})^{2}-(x'_{4})^{2}=x_{1}^{2}+x_{2}^{2}+x_{3}^{2}-x_{4}^{2}$,
so the transformation (\ref{eq:lt}) itself can be identified with a Lorentz
transformation (see also \cite{sexlurbantke:1992}, eq.~(2.1.7)), with $\eta=\alpha$
denoting the rapidity and relating the 4- (or 0-) coordinate to the standard
'time' coordinate. So $\cosh\eta=\gamma$, $\sinh\eta=\gamma\beta$ and $\tanh\eta=\beta$
with $\gamma=(1-\beta)^{-\frac{1}{2}}, \beta=\frac{v}{c}$. The same holds, of 
course, for the second point $y$.

Now the interesting and important fact is that $p_{34}$ (or $x_{3}y_{4}-x_{4}y_{3}$) 
by itself is invariant\footnote{In addition to direct calculation, one can 
use the $2\times 2$-determinant representation and standard rules of determinant 
calculus. This yields some more insight into standard rep theory if we 'transform' 
(or map) the determinant 'back' to the bilinear {\it point coordinate} 'invariant' 
\protect{$x^{T}y$} and use 'matrix' reps of group (or algebra) transformations,
i.e.~\protect{$y'=Gy, x'^{T}=(Gx)^{T}=x^{T}G^{T}$} with respect to the two columns.
Also note that if we complexify the coordinates, i.e.~\protect{$x_{3},x_{4},y_{3},y_{4}
\in\mathbbm{R}\longrightarrow\psi_{1},\psi_{2},\psi'_{1},\psi'_{2}\in\mathbbm{C}$},
we obtain expressions \protect{$\psi_{1}\psi'_{2}-\psi_{2}\psi'_{1}$} to be 
compared with (spinorial) singlet structures using the same mapping onto matrices
and e.g.~\protect{$\psi=\psi'^{+}$} with hermitean conjugation. This, however, will
be addresses in section {\ref{ch:liemapping}} in more detail.},
too, i.e. 
\begin{equation}
\begin{array}{rcl}
p'_{34} &=&x'_{3}y'_{4}-x'_{4}y'_{3} \\
&\stackrel{\protect{(\ref{eq:lt})}}{=}&
x_{3}y_{4}\,(\cosh^{2}\eta-\sinh^{2}\eta)+x_{4}y_{3}\,(\sinh^{2}\eta-\cosh^{2}\eta) \\
& = & x_{3}y_{4}-x_{4}y_{3} \\
&=&p_{34}\,.
\end{array}
\end{equation}
Without branching into the details of (Lorentz, Poincar\'{e} or general) group
theory as usual at this point (see e.g.~\cite{sexlurbantke:1992} or \cite{jackson:1983}),
for us it is sufficient having derived this property from a (linear) Complex
like $p_{12}\pm k p_{34}=0$. Here, we do not have room to start the geometrical
discussion of this Complex but we want to mention that this linear Complex
incorporates already special choices (see e.g.~\cite{pluecker:1865}, nrs.~28ff
or \cite{klein:1926}, \S\S 16, 17). Based mainly on \MO's and \PL's work 
(e.g.~\cite{moebius:1837}, \cite{pluecker:1865}, \cite{pluecker:1866}, 
\cite{plueckerNG:1868} and references therein), Klein recalls point-plane
mappings, null systems and line conjugation, and closes the gap to dynamics
and second order surfaces (i.e. polar theory).

Defining the axis of the null system as conjugate (line) polar of an absolute 
line and using a special choice of an orthogonal coordinate system, 
Klein discussed in detail how to obtain the rep 
$p_{12}\pm k p_{34}=0 \longrightarrow (xy'-x'y)+k(z-z')=0$,
$k$ denoting the 'parameter of the null system' via 
$k=\frac{a_{12}a_{34}+a_{13}a_{42}+a_{14}a_{23}}{a_{12}^{2}+a_{13}^{2}+a_{14}^{2}}$.
This notion is used as departure in various other contexts (see 
e.g.~\cite{ball:1876} or \cite{zindler:1902}) discussing aspects of 
(linear) line Complexe, sometimes using different notations. For our 
purpose, we use Lie's approach \cite{lie:1872} departing from the 
interpretation of imaginary elements in projective geometry and using 
them to establish a mapping of two complexified planar real coordinates
onto lines, i.e.~a 4-dim planar space to 4-dim line space.

\section{Lie Mapping (or Lie Transfer)}
\label{ch:liemapping}
While having the equivalent point/plane/line coordinate reps from
within \cite{dahm:MRST3}, section 1.4, in mind, we choose the line
rep for further discussion. In the general case, we thus have to 
discuss the (conjugation/reciprocity) relation of two linear Complexe 
(see e.g.~\cite{plueckerNG:1868}, \cite{klein:1869}, \cite{klein:1871}, 
\cite{reye:1866} II 18.~Vortrag, and especially \cite{lie:1872}, \S 7ff) 
versus occurences and use of quadratic Complexe. So Dirac's quest of 
finding a linear representation in $p$ (or sloppy speaking 'the root' 
of $p^2$) can be cast onto various related and interwoven topics:
\begin{enumerate}
\item[-] Given a quadratic Complex \textfrak{C}$^{2}$, how can we 
determine an appropriate linear Complex \textfrak{C}, its geometrical
and its dynamical and transformation/covariance properties?
\item[-] Given a quadratic Complex \textfrak{C}$^{2}$, can we find
other geometrical elements and/or representations \textfrak{C} which
square to a quadratic Complex, or what is the most general form of 
\textfrak{C}, respectively?
\item[-] Determining the parameters of (planar) line pencils or the
intersection points of lines e.g. with the absolute quadric, we obtain
quadratic relations in terms of line coordinates. How can we relate
them to physical observations and principles related to linear and 
quadratic line Complexe?
\end{enumerate}

\subsection{Complexification of 'Point' Coordinates}
% properties of line/Complex coordinates and their relation to
% 'quantum field theory' by complexification of point coordinates
Based on Lie's work\footnote{We use the notation 'Lie mapping' to 
denote the line-sphere mapping \protect{\cite{lie:1872}} which needs
to be distinguished from 'Lie transformations' used in more recent 
phase or function space discussions by means of Lie series resp.~power
series expansions. If we want to emphasize the background notion 
towards higher geometry over the more technical mapping aspects, 
we use also the notation 'Lie transfer'. This shouldn't signal that
especially reciprocity investigations or similar transfer mechanism
haven't been performed before in geometrical considerations, as one 
can easily check in literature. However, Lie's special considerations
provide a great contemporal overview and yield a beautiful application
at the same time.} in line and Complex geometry \cite{lie:1872}, \S 4, 
number 11, one can consider general reciprocity mappings between two 
real 3-dim spaces $r$ and $R$. For our purpose the specialization 
given in \S 7 eq.~(1), based on choosing two line Complexe, is 
substantial. So points in $r$ map to lines in $R$ and vice versa. 
Following \S\S 8 and 9, Lie obtained the special mapping\footnote{Standard 
small letters denote elements of the space \protect{$r$}, capital 
letters those of \protect{$R$}. Although $r$ is used parallel with 
\PL's line coordinate $r$, the respective context should suppress 
misunderstandings.} $2\rho=X+iY, 2s=X-iY, 2\sigma=Z\pm H, 2r=-(Z\mp H)=-Z\pm H$
which constrains the linear Complex $r+\sigma$ in $r$ mapped to 
minimal lines\footnote{Which later have to be expressed in Pauli or
spinorial counterparts.} in $R$ with $dX^{2}+dY^{2}+dZ^{2}=0$. 
Besides a lot of other features and results (see \cite{lie:1872} or
the overview in \cite{klein:1926} \S\S 25-27), this mapping accordingly 
transforms {\it lines} of the 3-dim real space $r$ uniquely into 
{\it spheres} of $R$. Vice versa, a sphere of $R$ described by $(X,Y,Z,H^{2})$, 
$\pm H=\sigma + r$, is transformed into only {\it two} lines of $r$!
Both lines $(X,Y,Z,+H)$ and $(X,Y,Z,-H)$ are polar with respect to
the linear Complex $\pm H=\sigma + r=0$ \cite{lie:1872}. The uniqueness
may be established by introducing the notion of orientation and Laguerre
geometry, however, according to our earlier footnote by introducing the 
matrix $M$, we can represent $M$ in terms of coordinates of $R$ by
\begin{equation}
\label{eq:quantum}
M\,=\,
\left(
\begin{array}{cc}
r & \rho \\ s & \sigma
\end{array}
\right)\quad\longrightarrow\quad
\left(
\begin{array}{cc}
\frac{1}{2}(-Z\pm H) & \frac{1}{2}(X+iY) \\ 
\frac{1}{2}(X-iY)    & \frac{1}{2}(Z\pm H)
\end{array}
\right)\,=\,\frac{1}{2}
\left(
\pm H\mathbbm{1}+X\sigma_{1}-Y\sigma_{2}-Z\sigma_{3}
\right)\,.
\end{equation}
i.e.~we obtain a mapping of lines (or line Complexe or even higher elements)
of the space $r$ into the space $R$. Lines of $r$ are mapped onto Pauli 
matrices $\sigma_{\alpha}$, $\sigma_{0}=\mathbbm{1}$, with real coefficients,
or simply SU(2). No need to recall the possibility to introduce quaternions 
here. However, it is apparent that higher line and Complex geometry have to 
find their appropriate counterparts in $R$ space if lines are mapped to Pauli 
matrices (or quaternions). So a priori we expect appropriate 'reflections' 
of higher line and Complex geometry in Clifford algebras.

It is however noteworthy, by starting from \PL's line equations $x=rz+\rho, y=sz+\sigma$
to mention one more fundamental issue where $x$ and $y$ denote (euclidean)
{\it projections} of a line in space (parametrized by $z$) onto the $xz$ 
and $yz$ planes, respectively. $\eta\sim (r y - s z) = r\sigma - s\rho = \det M$ 
denotes the projection onto the $xy$-plane\footnote{Please note once more the
{\it alternative} rep of the line by $x(t)$, $y(t)$ and $z(t)$ or by a 3-dim
euclidean 'vector' \protect{$\vec{x}(t)$} in terms of a time parameter and
a related velocity 'vector', time-dependent or not. Both pictures can be
'completed' by appropriate (additional) homogeneous coordinates, however,
it is necessary to care about the respective interpretations.}. If we rewrite
those equations in matrix rep, we obtain
\[
\left(
\begin{array}{c}
x\\y 
\end{array}
\right)\,=\,
\left(
\begin{array}{cc}
r & \rho \\ 
s & \sigma
\end{array}
\right)
\left(
\begin{array}{c}
z\\1
\end{array}
\right)
\quad\longrightarrow\quad
\left(
\begin{array}{c}
x_{1}\\x_{2} 
\end{array}
\right)\,=\,
\left(
\begin{array}{cc}
r & \rho \\ 
s & \sigma
\end{array}
\right)
\left(
\begin{array}{c}
x_{3}\\x_{4}
\end{array}
\right)
\]
after having introduced homogeneous (point) coordinates $x_{\alpha}$.
This requires to switch to quaternary matrix reps in order to describe
transformations of the full set of homogeneous (point) coordinates 
$x_{\alpha}$. It is obvious, that $2\times 2$ inverse and/or conjugate
matrices have to be considered when applying the Lie mapping, i.e.~we 
expect $2\times 2$ block structures in matrix reps of $R$ transformations.
Moreover, the $2\times 2$ calculus of the Lie mapping lightens some 
background of expressions like $\vec{\sigma}\cdot\vec{v}\vec{\sigma}\cdot\vec{v}=\vec{v}^{2}$
on the context of line coordinates and Complex conjugation and/or
quadratic Complexe.

At the time of writing, we tend to identify the space $r$ with physical
observable (projective) space whereas eq.~(\ref{eq:quantum}) represents
'quantum' reps respectively $R$ represents 'quantum space', e.g.~in terms
of $2\times 2$ Pauli or $4\times 4$ Dirac matrices considering two lines
(or Complexe). However, it is important to note that this relation yields
some assumptions related to linear (line) Complexe in $r$ \cite{lie:1872},
\S 7 and \S 8, and it is far from a general description especially if we 
relax the restrictions in $r$ and proceed to general dynamics and higher
geometry. Vice versa, we can establish a mapping to a Pauli/quaternionic
or Clifford calculus in order to relate physical observations and observables.

In \cite{lie:1872}, Lie presents various further very interesting results,
also with respect to Complex cones, differential geometry and mappings 
of tangent to curvature properties which we suppress here. However, there
are two more important facts which Lie used to introduce contact interactions
and his theory of partial differential equations but which for us is 
important to relate to physics. So he mentioned Klein's work on six 
Complexes in involution\footnote{Recall the handedness of the null systems!}
as a superset, in addition in \cite{lie:1872}, \S 9, number 28, he showed
that two incident lines (being reciprocal/conjugate polars with respect
to the special Complex $H=0$) are mapped to two spheres being in contact, 
i.e.~fulfilling $(X_{1}-X_{2})^{2}+(Y_{1}-Y_{2})^{2}+(Z_{1}-Z_{2})^{2}=(H_{1}\pm H_{2})^{2}$.
If we rewrite the (line) incidence relation like above (see also 
\cite{dahm:MRST3} section 2.1) and respect \cite{plueckerNG:1868},
\cite{klein:1869}, \cite{klein:1871}, the action can be treated by
quadratic Complex theory. There is an interesting side effect when
restricting observations to planar problems in that we may investigate
(planar) Complex curves and still discuss (energy) conic sections in
the planes as well as two pencils of lines related to two linear Complexe
(and their congruence). We'll address this elsewhere.

%\subsection{Compact Groups and Spinor Representations}
% SU(2)
% SU(2)$\times$U(1)
% Dirac spinor description and plane waves
%Lie's sphere geometry <-> elliptic geometry and quaternions

\subsection{Energy and Second Order Surfaces}
\label{ch:energy}
% definition of energy, addition of 'gauge terms' by parameters/constraints
Last not least it is noteworthy to focus on the description of physical
actions which typically starts from an action principle and variations.
From above, we are already equipped with an action principle; from the
background of projective geometry, we are equipped with a coordinate
tetrahedron, with transformation groups and (linear) representation
theory, and last not least, we may use transfer principles.

% power expressions, compare to six-vector products of Complexe
So from above, we may include (invariant) products like $F^{\mu\nu}F_{\mu\nu}$
or $F^{a\,\mu\nu}F^{a}_{\mu\nu}$, $1\leq a\leq 3$, and Complex geometry.
However, we may as well try to find appropriate representations of line
Complexe on other rep spaces. As such, we can observe that the product
of two (real) quaternions $a^{+}, b$, $^{+}$ denoting quaternionic 
conjugation, yields Complex coefficients if we neglect the trace for
the moment. So the mapping $a^{+}\cdot b\longrightarrow A_{\alpha\beta}\, 
q^{+}_{\alpha}q_{\beta}\longrightarrow A_{\alpha\beta}\,q_{\alpha}\otimes q_{\beta}$ 
allows to extract the relevant (real) coefficients and relate them to 
SL(2,$\mathbbm{H}$) and its real forms. On the other hand, placing two
quaternions $q^{+}, q$ into a 'spinor', we find of course a rotated
'spinor' set $1/2(q^{+}\pm q)$ which can be related to Dirac theory
and Clifford algebra. In the background, it is of course the Lie 
mapping which is active in relating lines and Complexe to spheres
and sphere Complexe, etc. For our purpose here, however, it is 
sufficient to justify the 'spinorial rep' of Complexe in an action
principle which is based on the Lie mapping. For all calculations,
the standard mechanisms can be applied to calculate energies and
states, however, the interpretation has to be changed at a couple
of places. We are going to investigate such topics in upcoming 
work.

\end{document}